\documentclass[aps,prl,floatfix,amsmath,dstroke,amssymb,prlsymb,twocolumn, showpacs,superscriptaddress]{revtex4}% You should use BibTeX and apsrev.bst for 
\usepackage{graphicx}
\usepackage{dsfont}
\begin{document}

\title{Extreme statistics of complex random and quantum chaotic states}
\author{ Arul Lakshminarayan \footnote{Permanent address: Department of Physics, Indian Institute of Technology Madras, Chennai, 600036, India.}}
\author{Steven Tomsovic \footnote{Martin Gutzwiller Fellow, 2006/2007; permanent address: Department of Physics and Astronomy, Washington State University, Pullman, WA  99164-2814}}
\affiliation{Max-Planck-Institut f\"ur Physik komplexer Systeme, N\"othnitzer Stra$\beta$e 38, D-01187 Dresden, Germany}
\author{Oriol Bohigas}
\author{ Satya N.~Majumdar}
\affiliation{Laboratoire de Physique Th\'eorique et Mod\`eles Statistiques (UMR 8626 du CNRS), Universite Paris-Sud, B\^atiment 100, 91405 Orsay Cedex, France.}
%\affiliation{Department of Physics\\ Indian Institute of Technology Madras\\
%Chennai, 600036, India.}
%\author{Krishnendu Maity}
%\email[]{ kmaity@gmail.com}
%\author{Arul Lakshminarayan}
%\email[]{arul@physics.iitm.ac.in}
%\affiliation{Department of Physics\\ Indian Institute of Technology Madras\\
%Chennai, 600036, India.}

%\preprint{IITM/PH/TH/2006/3}

%\date{\today}
\begin{abstract}
An exact analytical description of extreme intensity statistics  in complex random states 
is derived.  These states have the statistical properties of the Gaussian and Circular Unitary Ensemble eigenstates of random matrix theory.   Although the components are correlated by the normalization constraint, it is still possible to derive compact formulae for all values of the dimensionality $N$.  The maximum intensity result slowly approaches the Gumbel distribution even though the variables are bounded, whereas the minimum intensity result rapidly approaches the Weibull distribution. Since random matrix theory is conjectured to be applicable to chaotic quantum systems, we calculate the extreme eigenfunction statistics for the  standard map with parameters at which its classical map is fully chaotic.  The statistical behaviors are consistent with the finite-$N$ formulae.

\end{abstract}
\maketitle

%\newpage

\newcommand{\newc}{\newcommand}
\newc{\beq}{\begin{equation}}
\newc{\eeq}{\end{equation}}
\newc{\kt}{\rangle}
\newc{\br}{\langle}
\newc{\beqa}{\begin{eqnarray}}
\newc{\eeqa}{\end{eqnarray}}
\newc{\pr}{\prime}
\newc{\longra}{\longrightarrow}
\newc{\ot}{\otimes}
\newc{\rarrow}{\rightarrow}
\newc{\h}{\hat}
\newc{\bom}{\boldmath}
\newc{\btd}{\bigtriangledown}
\newc{\al}{\alpha}
\newc{\be}{\beta}
\newc{\ld}{\lambda}
\newc{\sg}{\sigma}
\newc{\p}{\psi}
\newc{\eps}{\epsilon}
\newc{\om}{\omega}
\newc{\mb}{\mbox}
\newc{\tm}{\times}
\newc{\hu}{\hat{u}}
\newc{\hv}{\hat{v}}
\newc{\hk}{\hat{K}}
\newc{\ra}{\rightarrow}
\newc{\non}{\nonumber}
\newc{\ul}{\underline}
\newc{\hs}{\hspace}
\newc{\longla}{\longleftarrow}
\newc{\ts}{\textstyle}
\newc{\f}{\frac}
\newc{\df}{\dfrac}
\newc{\ovl}{\overline}
\newc{\bc}{\begin{center}}
\newc{\ec}{\end{center}}
\newc{\dg}{\dagger}
\newc{\prh}{\mbox{PR}_H}
\newc{\prq}{\mbox{PR}_q}

The study of the statistics of extreme values \cite{Gumbel04}  has  found many applications in diverse areas 
such as geophysics, metereology, economics, structural engineering, ocean waves, and dynamical
systems.  The subject is currently undergoing a resurgence of interest due to recent catastrophic events such as hurricanes, floods, and a particularly deadly tsunami as well as a number of research advances \cite{Albeverio06}. The kinds of questions being asked are, for example, what are the distributions for extreme events, or what are the inter-event gap distributions. It has long been known that if the underlying events are independent and identically distributed, then for appropriately rescaled variables there are three possible limiting universal distributions for the extreme maximal events:  the Fr\'echet, Gumbel and Weibull distributions \cite{Gumbel04}.   Respectively, they arise depending on whether the tail of the density is a power law, or faster than any power-law, and unbounded or bounded.  If there are correlations, then it is known that these universal distributions are reached for a sufficiently fast  decay of auto-correlations \cite{Leadbetter88}.

In this letter, we apply these powerful methods to study the extreme properties of  random vectors, or wave functions more generally, in the context of quantum mechanics.
Our motivation is that the eigenstate intensities in fully chaotic systems with no particular symmetries are conjectured to behave exactly as these random vectors subject only to a normalization constraint as in random matrix theory.  For chaotic systems, the applicability of  random matrix theory \cite{Mehta04,Brody81} has been well appreciated since the Bohigas-Giannoni-Schmit conjecture which states that the spectral fluctuations of quantized classically chaotic systems can be modelled by a suitable  ensemble of random matrices \cite{Bohigas84}.  In fact, a certain number of extreme spectral properties have already been derived \cite{Majumdar03, Gyorgyi03, Dean06, Sabhapandit07}  beginning with the well-known result for the distribution of the largest eigenvalue \cite{Tracy94, Tracy96}.  The corresponding treatment of random vectors or quantum eigenvectors has not yet been addressed. However, see \cite{Aurich99} for an initial foray into random waves.

In fact, eigenstate intensities in strongly chaotic systems are known to follow an exponential density, which is consistent with states uniformly distributed over a standard simplex \cite{Bengtsson06}, as happens in the Unitary Ensembles  (if an anti-unitary symmetry is respected, Porter-Thomas density, hypersphere, and Orthogonal Ensembles \cite{Porterbook}).  A similar class of problems shows up in fragmentation, i.e.~randomly cutting an object  of fixed length into $N$ pieces \cite{Derrida87}.  

It is possible to give compact, exact formulae for all dimensionality $N$ in spite of the correlations introduced by the normalization constraint.  It turns out that the small-$N$ distributions for the maxima differ considerably from their asymptotic limit (which turns out to be Gumbel, $\exp[-{\rm e}^{-x}]$) thus giving the possibility of extracting system size from the distributions.  In an $N$-dimensional complex Hilbert space a general state is represented in a
fixed orthonormal basis  $|i\kt$ as $|\psi\kt = \sum_{i=1}^{N} z_i |i \kt$.  If the $z_i$ are  complex components of a random state, they are not constrained by any requirement other than normalization, and their joint probability distribution is:
\beq
P(z_1,z_2,\ldots,z_N)=\df{(N-1)!}{\pi^N}\delta\left(\sum_{j=1}^N|z_j|^2\, -\, 1\right).
\eeq
The real and imaginary parts of the components are spread in an unbiased, microcanonical, manner over the $2N$-dimensional unit sphere. The intensities $|z_i|^2$ are distributed uniformly on an $N-1$ simplex.  Thus a random state is correlated by this requirement.  Given such an ensemble of random states we consider the probability distribution $\rho(t,N)$ of  $t=\mbox{max}\{|z_1|^2,|z_2|^2,\cdots,|z_N|^2\}$.  Let $F(t,N)$ be  the probability that {\em all} $|z_j|^2 \le t$. This is called the distribution or cumulative density of such extreme events, {\it i.e.} $F'(t,N)=\rho(t,N)$ where the prime denotes
differentiation with $t$. Thus 
\beq
\label{maxint}
F(t,N)=\df{(N-1)!}{(2\pi)^N}\prod^N_{i=1}\left[ \int_{0}^{2 \pi} d\theta_i\int_{0}^t  dr_i^2\right] \delta\left(\sum_{j=1}^Nr_j^2\, -\, 1\right)
\eeq
where $r_j\equiv |z_j|$.
Use a Fourier decomposition of the $\delta$-function and perform the $\{\theta_i,r_i \}$ integrals to arrive at
\beq
F(t,N)=\df{ (N-1)!}{2\pi }\int_{-\infty}^{\infty} d \xi e^{-i \xi} \left(\df{e^{i \xi t} -1}{i \xi} \right)^N.
\eeq
Next expand the power in a binomial series to find
\beq
F(t,N)= \df{ (N-1)!}{2\pi i^N}  \sum_{m=0}^{N} \binom{N}{m}  (-1)^{N-m} I(1-mt),
\eeq
where
\beq
I(\omega)=\int_{-\infty}^{\infty} e^{-i \omega \xi } \xi^{-N} d \xi= \df{\pi (-i)^N }{(N-1)!} \omega^{N-1}\, \mbox{sign}(\omega).
\eeq
The exact result for $F(t,N)$ is
\beq
\label{exactsun}
F(t,N)= \df{1}{2}  \sum_{m=0}^{N} \binom{N}{m}  (-1)^{m} (1-mt)^{N-1} \mbox{sign}(1-mt).
\eeq 
Expanding the power again and using the identity \cite{Gradshteyn00}:
\beq
\sum_{m=0}^N \binom{N}{m} (-1)^m m^k =0,\;\; k=0,\ldots N-1
\eeq
gives ``resummed'' expressions for $F(t,N)$ valid in the intervals $I_k=[1/(k+1),1/k]$, where $k=1,2,\cdots,N-1$:
\beq
\label{finformsun}
F( t \in I_k ,N) = \sum_{m=0}^k \binom{N}{m} (-1)^m (1-mt)^{N-1},
\eeq
and  $F(t\le 1/N,N)=0$.  Thus the cumulative density is a piecewise smooth function on the intervals $I_k$. 

Given the simple form of the distribution above, it is useful to interpret them combinatorially and 
derive them from such an approach.  First note that
\beq
P_l(z_1,z_2,\ldots,z_l)=\df{\Gamma(N)}{\pi^l \Gamma(N-l)} \left( 1-\sum_{j=1}^{l} |z_j|^2 \right)^{N-l-1}
\eeq
where $P_l$ is the reduced probability density for $l$ complex components ($P_N$ is identical to the
full distribution $P$ above).  If  $t \in I_1=[1/2,1]$ there can be at most only one such component. Therefore, the fraction of states with a component larger than $t$ is exactly the fraction of components larger than $t$. Since the desired quantity is the fraction of states such that all components are less than $t$, it is the simply the complement:
\beq
F(t,N)=1-N \int_{|z|^2\ge t} P_1(z) dz
\eeq
The factor $N$ accounts for multiplicity of choice of this one component.
This integral is elementary for the complex case and gives $F(t,N)=1-N (1-t)^{N-1}$, which agrees with 
the series in the RHS of Eq.~(\ref{finformsun}) which terminates at $k=1$ for $t\in I_1$.

If $t\in I_2$, it is possible that there are {\it at most} two such components.  The number of components $\ge t$ is no longer the number of sequences (states) with at least one component more than $t$ as it double counts states which have two components larger than $t$, a contribution which must get subtracted.  This same logic extends, and in the next interval $I_3$, the number of  pairs over counts the contributions of triples.  Similarly, this reasoning carries forward to any distribution with a {\it unit norm constraint} and gives for the cumulative density  
\begin{eqnarray}
\label{general}
&&F(t \in I_k,N) =  \nonumber \\
&&\sum_{m=0}^{k} \binom{N}{m} (-1)^m \int_{|x_i|^2>t} P_m(x_1,\ldots,x_m)\, dx_1\cdots dx_m, \nonumber \\
\end{eqnarray}
which is the generalization of Eq.~(\ref{finformsun})  (the $m=0$ term in the 
above expression is unity).  This generalization could be used as the starting point for an analysis of real random states as well as general density matrix eigenvalues whose sum is also constrained to be unity.  It is interesting that piecewise continuous extreme distributions have been identified in dynamical systems \cite{Balakrishnan95,Nicolis06} and fragmentation problems \cite{Derrida87}, where a similar combinatorial approach was also applied.

From the distribution Eq.~(\ref{exactsun}) above, it is possible to derive exact formulae for the moments.  In particular, results for the first (mean) and the second moments of the maximum components are
\beqa
\label{moments}
\br t \kt& =& \df{H(N,1)}{N} = \df{\gamma + \ln N}{N}+{\mathcal O}\left( \df{1}{N^2}\right) \\
\br t^2\kt &=& \df{H^2(N,1)+H(N,2)}{N(N+1)}
\eeqa
where $H(N,k)$ is the Harmonic number of order $k$ defined by the finite sum $\sum_{m=1}^N m^{-k}$,
and $\gamma=0.5772\cdots$ is the Euler-Mascheroni constant. The standard deviation can be calculated exactly and its large-$N$ form is: 
\beq
\label{sigma}
\sigma(t)=\df{\pi}{\sqrt{6} N} + {\mathcal O} \left(\df{\ln(N)}{N}\right)^2.
\eeq
It turns out that the asymptotic $N\rarrow \infty$ limit here is also the result for uncorrelated exponentially distributed variables of mean $1/N$. This limit may be calculated by simply taking $P_1$ as the probability density of an independent process, which gives
\beq
\label{ftni}
F(t,N\rightarrow\infty)=\left(1-e^{-Nt}\right)^N \approx\exp\left(-e^{-N[t-\ln N /N]}\right).
\eeq
Expressed in terms of the standard linearly scaled variable $x=(t-a_N)/b_N$, this distribution is seen to coincide with the Gumbel distribution where the parameters are given by $a_N=\ln(N)/N $ and $b_N=1/N$. As should happen, the mean and the standard deviation calculated from the Gumbel distribution coincide with the leading order contributions derived in Eqs.(\ref{moments},\ref{sigma}).
\begin{figure}
\includegraphics[height=2.4in,angle=0]{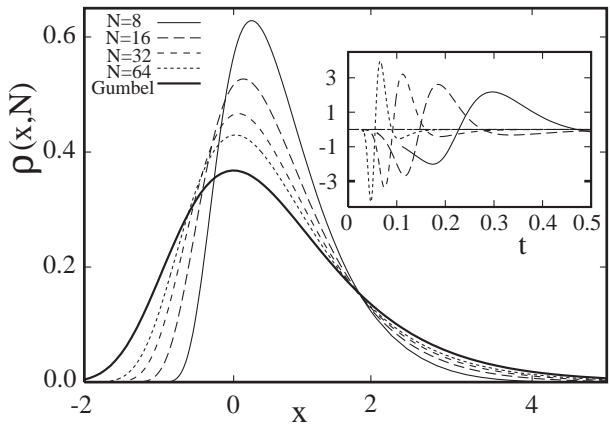}
\caption{The convergence of the exact probability density  to the asymptotic Gumbel distribution using the scaled variable $x=N(t-\ln(N)/N)$ with increasing $N$. The inset shows the difference between the exact and the Gumbel distribution for the same values of $N$, but in the unscaled variable.}
\label{gumb}
\end{figure}  
That the maximal statistics is a Gumbel distribution is interesting because the intensities have a finite support, being constrained to lie on the standard $N-1$-simplex. The expectation for uncorrelated bounded variables would be the Weibull distribution; however the correlations introduced by the normalization constraint transforms this to the Gumbel distribution.  In Fig.~(\ref{gumb}) we compare the exact probability density with that of the Gumbel density ($\exp[-x-e^{-x}]$) after appropriately rescaling. It is clear that the approach of the exact density to the Gumbel one is rather slow and at $N$ around $100$ there are still significant differences between the asymptotic and the exact. It is also instructive to note that without the rescaling the densities actually diverge as $N$ increases.  This difference gives the possibility of extracting system size information from the extreme statistics.

The distribution of the {\it minimum} intensity $s=\mbox{min}\{|z_1|^2,|z_2|^2,\cdots,|z_N|^2\}$
on the other hand is a much simpler quantity and is not asymptotically a Gumbel distribution.
The fraction of states such that the minimum is larger than some $s$ is the fraction of states such
that all the components are larger than $s$. Thus if $F(s,N)$ is the cumulative distribution of the
minimum, it is given by 
\beq
\label{minint}
F(s,N)=1-\df{(N-1)!}{(2\pi)^N}\prod_{i=1}^N \left[\int_{0}^{2 \pi} d\theta_i\int_s^{1} dr_i^2\right] \delta\left(\sum_{j=1}^Nr_j^2\, -\, 1\right),
\eeq
which is very similar to the integral in Eq.~(\ref{maxint}). Its evaluation proceeds similarly to the maximum, and gives:
\beq
\label{mincum}
F(s,N)=\left\{\begin{array}{ll}1-(1-Ns)^{N-1} &0\le s \le 1/N\\
1 &1/N \le s \le 1 \end{array}\right..
\eeq
It is clear that the minimum cannot exceed $1/N$, just as the maximum cannot be less than this.
The average minimum component is easily calculated and is 
exactly equal to $\br s\kt =1/N^2$.
The distribution for the minimum does not have the piecewise continuous character we observed
for the maximum. This has a geometrical interpretation in terms of the standard-simplex. In the case of the maximum
the integral in Eq.~(\ref{maxint}) can be interpreted in terms of volumes of subsets contained in the region bounded 
by the standard $N-1$-simplex; these volumes enclosing more complex shapes for increasing $t$, as they
pierce the simplex boundary.  Whereas for the integral in Eq.~(\ref{minint}) the volumes involved are those of the entire simplex and the volume of a subset that never pierces the $N-1$ simplex. 

For large $N$ the distribution of the minimum approaches the exponential:
\beq
F(s,N\rightarrow\infty)=1-\exp(-N^2 s).
\eeq
This being a special case of the Weibull distribution, it is indeed what one would expect of 
uncorrelated variables with a compact support. That the minimum cannot be less than zero
presents a strong constraint and for small components the normalization correlation is not so important. Thus, the distribution of the maximum and minimum of the complex random vector intensities follow different universal distributions asymptotically.  It is noteworthy, however, that the limiting large deviations of the maximum component toward a small value which occur in the interval $I_{N-1}=1/N \le t \le 1/(N-1)$ is distributed as $F(t,N)=(Nt-1)^{N-1}$ which is an exact reflection about the value $1/N$ of the behavior of the minimum component.  

In order to compare these extreme statistics to the statistical properties of the eigenfunctions of a Hamiltonian system, consider a quantum kicked rotor \cite{Izrailev90}. This is a stroboscopic mapping of a kicked one-dimensional particle of unit mass moving on  a circle of unit perimeter with the Hamiltonian $H(q,p) = p^2/2 -(K/4\pi^2)\cos(2 \pi q) \sum_{n=-\infty}^\infty \delta(t-n)$.   From $H(q,p)$, the classical mapping can be given \cite{Lichtenberg83} (modulo unity):
\begin{equation}
\label{kreq}
p_{i+1} =  p_i -K/(2\pi)\sin(2 \pi q_i)\ ;  \quad q_{i+1} = q_i + p_{i+1} 
\end{equation}
where the potential kick is applied before the free motion and the phase space is restricted to the unit torus ($q,p\in [0,1)$) for convenience. For $K \gg 5$ the map is highly chaotic, although the phase space almost always contains some tiny proportion of regular motion mixed in. 

In the position basis, the evolution operator is 
\beq
\label{qmap}
\begin{split}
U_{n n'}&= \f{1}{N}\exp\left(\frac{i N K}{2\pi}\cos\left(\f{2\pi (n'+\al)}{N}\right)\right)\\  & \times \sum_{m=0}^{N-1}
\exp\left(- \pi i \f{(m+\be)^2}{N} + \f{2\pi i (m+\be)(n-n')}{N} \right).
\end{split}
\eeq
The two phases $0 \le \al,\be \le1$ can be used for controlling parity and time-reversal symmetry breaking ($\al=1/2,\be=0$ preserves both symmetries).  Choosing $\be$ well away from $0$ or $1/2$ breaks time-reversal invariance, which for $K\gg 5$ leads to quantum chaotic states that are complex  and whose extremes should follow the distributions of the unitary ensembles.  
\begin{figure}
\includegraphics[height=2.5in,angle=0]{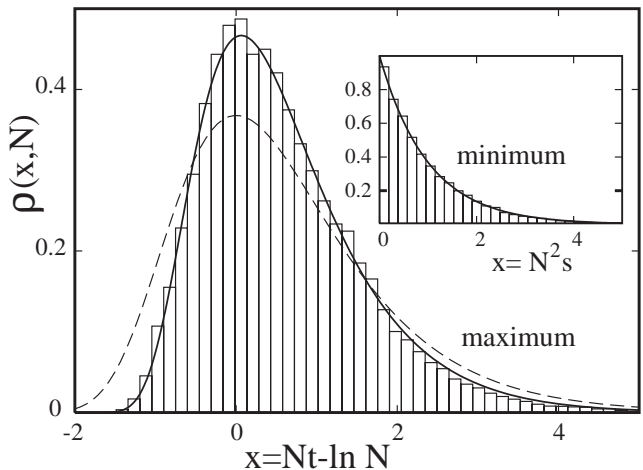}
\caption{The probability densities (histograms) of the scaled maximum and minimum (inset) intensity of eigenfunctions in the position basis of the quantum kicked rotor for $N=32$ in the parameter range $13.8<K<14.8$.  Shown as a continuous line is the exact density for random states while the dotted ones are the respective Gumbel and Weibull densities. }
\label{fig2}
\end{figure}  
The dimensionality of the Hilbert space $N$ is the inverse (scaled) Planck constant.  

An ensemble of roughly $30,000$ quantum chaotic eigenstates is created from the $N$ orthonormal states, as well as from those obtained by a variation of parameters $K$, $\al$, and $\be$ such that while the quantum spectrum is significantly changed, the classical dynamics is not.  Figure~(\ref{fig2}) shows the density of the maximal and minimal intensities in the position basis for a range of $K$ values where the classical map has no significant islands and is highly chaotic \cite{Tomsovic07}.  The derivatives of the exact results in Eqs.~(\ref{finformsun},\ref{mincum}) fit the quantum system histograms very well. The deviations are roughly of the scale of the expected sample size errors.  Note that the dynamical system results require the exact density for the maximum as the asymptotic approach is too slow.  This is in contrast to the exact finite-$N$ Dyson/Mehta fluctuation measures \cite{Mehta04}, which approach their asymptotic limits so quickly that finite-$N$ results are rarely, if ever, used.  The inset shows that the fit to the Weibull is excellent even for the small value of $N$ used.  Given the excellent agreement of the dynamical systems results with the analytic forms, deviations may in turn be used to investigate the important issue of eigenstate localization. 

In summary, recent years have seen a fast and fruitful development of the merging of extreme statistics tools and random matrix eigenvalues.  This letter extends this general perspective to the properties of eigenvectors.

\bibliography{classicalchaos,extreme,general_ref,quantumchaos,rmt}

\end{document}